\documentclass[]{spie}  

\usepackage{amsmath,amsfonts,amssymb}
\usepackage{graphicx}
\usepackage[colorlinks=true, allcolors=blue]{hyperref}

\title{Simons Observatory large aperture receiver simulation overview}
\author[a]{John L. Orlowski-Scherer}
\author[a]{Ningfeng Zhu}
\author[a]{Zhilei Xu}
\author[b]{Aamir Ali}
\author[c]{Kam S. Arnold}
\author[b]{Peter C. Ashton}
\author[a]{Gabriele Coppi}
\author[a]{Mark Devlin}
\author[a]{Simon Dicker}
\author[c]{Nicholas Galitzki}
\author[d]{Patricio A. Gallardo}
\author[c]{Brian Keating}
\author[b,e,f]{Adrian T. Lee}
\author[a]{Michele Limon}
\author[g]{Marius Lungu}
\author[h]{Andrew May}
\author[i]{Jeff McMahon}
\author[d]{Michael D. Niemack}
\author[h]{Lucio Piccirillo}
\author[j]{Giuseppe Puglisi}
\author[k]{Maria Salatino}
\author[e]{Max Silva-Feaver}
\author[i]{Sara M. Simon}
\author[a,l]{Robert Thornton}
\author[h]{Eve M. Vavagiakis}

\affil[a]{University of Pennsylvania, 209 South $33^{\text{rd}}$ St., Philadelphia, United States}
\affil[b]{Department of Physics, University of California, Berkeley, Berkeley, CA, USA}
\affil[c]{Department of Physics, University of California San Diego, La Jolla, CA, USA}
\affil[d]{Department of Physics, Cornell University, Ithaca, NY USA}
\affil[e]{Physics Division, Lawrence Berkeley National Laboratory, Berkeley, USA}
\affil[f]{Radio Astronomy Laboratory, University of California, Berkeley, Berkeley, CA, USA}
\affil[g]{Department of Physics, Princeton University, Princeton, NJ, USA}
\affil[h]{School of Physics \& Astronomy, University of Manchester, Manchester, UK}
\affil[i]{Department of Physics, University of Michigan, Ann Arbor, MI, USA}
\affil[j]{Department of Physics, Stanford University, Stanford, California, CA, USA}
\affil[k]{AstroParticle and Cosmology Laboratory, Paris Diderot University, Paris, France}
\affil[l]{Department of Physics \& Engineering, West Chester University of Pennsylvania, West Chester, PA, USA}

\authorinfo{Further author information: Send correspondence to John Orlowski-Scherer, E-mail: jorlo@sas.upenn.edu, Telephone: 1 732 236 0914}

\begin{document} 
\maketitle

\begin{abstract}
The Simons Observatory (SO) will make precision temperature and polarization measurements of the cosmic microwave background (CMB) using a series of telescopes which will cover angular scales between one arcminute and tens of degrees, contain over 60,000 detectors, and sample frequencies between 27 and 270 GHz. SO will consist of a six-meter-aperture telescope coupled to over 30,000 detectors along with an array of half-meter aperture refractive cameras, which together couple to an additional 30,000+ detectors. SO will measure fundamental cosmological parameters of our universe, find high redshift clusters via the Sunyaev-Zeldovich effect, constrain properties of neutrinos, and seek signatures of dark matter through gravitational lensing. In this paper we will present results of the simulations of the SO large aperture telescope receiver (LATR). We will show details of simulations performed to ensure the structural integrity and thermal performance of our receiver, as well as will present the results of finite element analyses (FEA) of designs for the structural support system. Additionally, a full thermal model for the LATR will be described. The model will be used to ensure we meet our design requirements. Finally, we will present the results of FEA used to identify the primary vibrational modes, and planned methods for suppressing these modes. Design solutions to each of these problems that have been informed by simulation will be presented. 
\end{abstract}


\section{INTRODUCTION}
The cosmic microwave background (CMB) has become one of the most powerful probes of the early universe. Measurements of temperature anisotropies on the level of approximately ten parts per million have brought cosmology into a precision era, and have placed tight constraints on the fundamental properties of the universe. Beyond temperature anisotropies, CMB polarization anisotropies not only enrich our understanding of our cosmological model, but could potentially provide clues to the very beginning of the universe via the detection (or non-detection) of primordial gravitational waves.  A number of experiments have made and are continuing to refine measurements of the polarization anisotropy.  However, these experiments are typically dedicated to a relatively restricted range of angular scales, e.g., large angular scales (tens of degrees) or high resolution/small angular scales (on the order of 1\,arcminute). To provide a complete picture of cosmology, both large and small angular scales are important. Ideally these measurements would be made from the same observing site so that the widest range of angular scales can be probed, at multiple frequencies, on the same regions of the sky. This is the goal of the Simons Observatory (SO).  SO will field a 6-meter large aperture telescope (LAT) coupled to the large aperture telescope receiver. During initial deployment, seven of the planned thirteen optics tubes will be installed in the LATR, containing over 30,000 detectors. The LAT is designed with a large FOV capable of supporting a cryostat with up to 19 LATR-like optics tubes.  To limit the development risk of the large SO cryostat, the LATR is designed to accommodate up to 13 optics tubes. We plan to deploy 7 optics tubes with 3 detector wafers in each for a total of roughly 35,000 detectors, primarily at 90/150~GHz in the initial SO deployment.  We note that each optics tube could be upgraded to support 4.5 wafers for a ~50\% increase in the number of detectors per optics tube. With this upgrade and the deployment of 19 optics tubes, the LAT could support roughly 145,000 detectors at 90/150~GHz.
 SO will also have an array of half-meter large angular scale cameras coupled to an additional 30,000 detectors. The unique combination of telescopes in a single CMB observatory, which will be located in Chile\textsc{\char13}s Atacama Desert at an altitude of 5190~m, will allow us to sample a wide range of angular scales over a common survey area. 

In this paper we will provide an overview of the simulations done in support of the LATR design process. Finite element analysis (FEA) simulations give us critical feedback about the performance of our components, which we use to refine their design. In Sec.~\ref{sec:mech} we cover the suite of mechanical simulations that we did in support of our design using the Solidworks Simulation module.\footnote{Dassault Syst\`emes, 10, Rue Marcel Dassault, 78140, V\'elizy-Villacoublay, FRANCE, https://www.solidworks.com/} Then, in Sec.~\ref{sec:therm} we described our combined thermal model and detail the thermal gradient simulations performed with the COMSOL software.\footnote{COMSOL, Inc., 100 District Avenue, Burlington, MA 01803, USA, https://www.comsol.com}
The challenges we faced when designing the LATR are unique in that we are trying build the largest ground-based CMB receiver to date. However, the solutions to these challenges we developed with the assistance of FEA simulations will provide a critical stepping stone for the next generation CMB experiments, in particular, CMB-S4\cite{Abazajian2016,Abitbol2017}.

\section{MECHANICAL Simulations}
\label{sec:mech}
In order to ensure the structural stability of our receiver, a full suite of FEA simulations of all structurally critical components was performed. All structural simulations were done using the Solidworks Simulation software module. The core result in each simulation is the static simulation, which quantifies the linear strain of our receiver under static conditions. Our criterion for a success is if the minimum factor of safety (FoS) is four or higher, where the FoS is the elastic yield strength divided by the predicted strain. Additionally, we perform a buckling analysis. The buckling analysis examines the behavior of thin elements under compression, and quantifies whether those parts will collapse due to loss of stability. The criterion for a successful buckling simulation is a load factor greater than 4, where the load factor is the strain at buckling divided by the maximum predicted strain, analogous to FoS. Finally, for some temperature stages, we perform a vibrational analysis, which determines the fundamental frequencies of vibrations for our various temperature stages. Vibrations present a number of challenges: they can put the optics out of alignment, they work harden the copper we use to conduct heat, thereby reducing its efficacy, and, in our coldest stages, can cause appreciable heating. The G10 that we use to support our 80~K, 40~K, and 4~K stages is very stiff, and thus acts as a high frequency filter on vibrational modes coming from the 300~K stage. In particular, we require that the first vibrational mode of each stage is several times higher than the first vibrational mode of the telescope, at 3~Hz, so that the vibrations of the telescope do not couple to the receiver. Therefore, we consider our vibrational analysis a success if the first fundamental mode is at 20~Hz or higher.\\

For all simulations, the key features are loads, fixtures, contacts, and material properties. In all simulations, we include the gravitational load from the masses involved. In any simulations where there is a mass load but the corresponding body is not in the solid model, we use a displaced load to simulate the force. An example is the 4~K stage, where the 1~K and 100~mK bus masses are suspended from the 4~K plate, but the buses themselves are not included in the simulation for simplicity. Each simulation has a fixture, a surface which is defined not to move or deform under load. Generally this is the flange from which the stage is supported. For example, in the 40~K simulation, the flange from which it is supported is the fixture. While these fixtures can deform slightly in reality, we have done simulations of the full structural path which confirm that the flanges deform minimally. Component contacts define the interface between parts. In general, we assume that parts are bonded, which causes our results to slightly over estimate the safety of our designs, and this motivates the high FoS. Finally, material properties are the defaults from Solidworks, with two exceptions. The two major materials which we defined custom materials for are G10\cite{Kasen1980}, and carbon fiber, where we obtained technical specification from the planned vendor, vanDjik Pultruded Products.\footnote{Aphroditestraat 24, NL-5047, TW TILBURG, The Netherlands, http://www.dpp-pultrusion.com/en/the-company/}

\subsection{Vacuum Shell}
\label{sec:shell}
\begin{figure} [ht]
   \begin{center}
   \begin{tabular}{c} 
   \includegraphics[width = .75\linewidth]{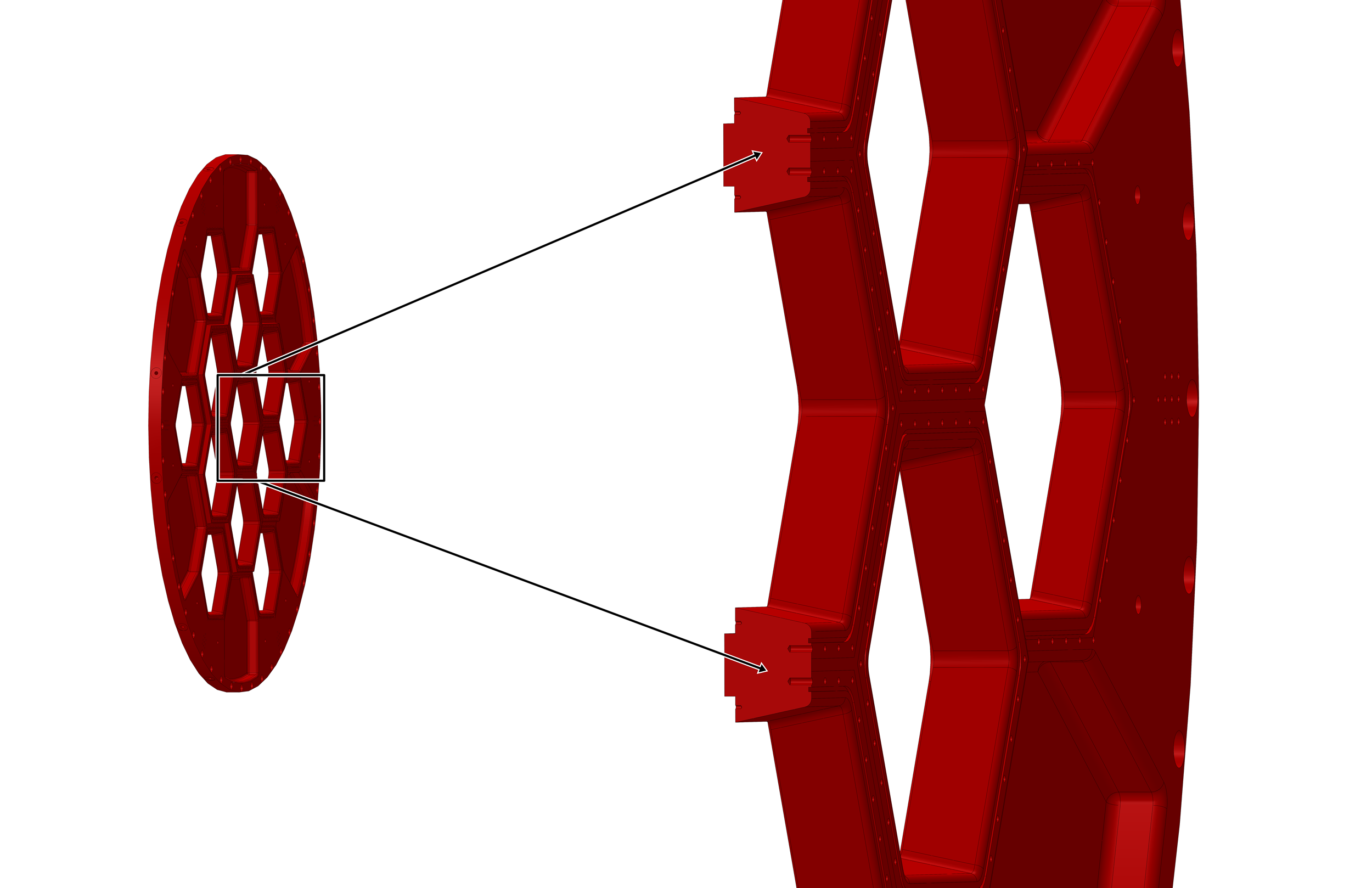}
   \end{tabular}
   \end{center}
   \caption[example] 
   { \label{fig:Taper} 
Cross section of front plate showing the tapering of the windows with the beam convergence. Light is entering the cryostat from the right, and detector arrays are to the left. Since we tapper the windows, the left hand side of the front plate is thicker than it would be if we had cut the entire window to the beam size at the right hand side of the window.}
   \end{figure}

\begin{figure} [ht]
   \begin{center}
   \begin{tabular}{c} 
   \includegraphics[width = .75\linewidth]{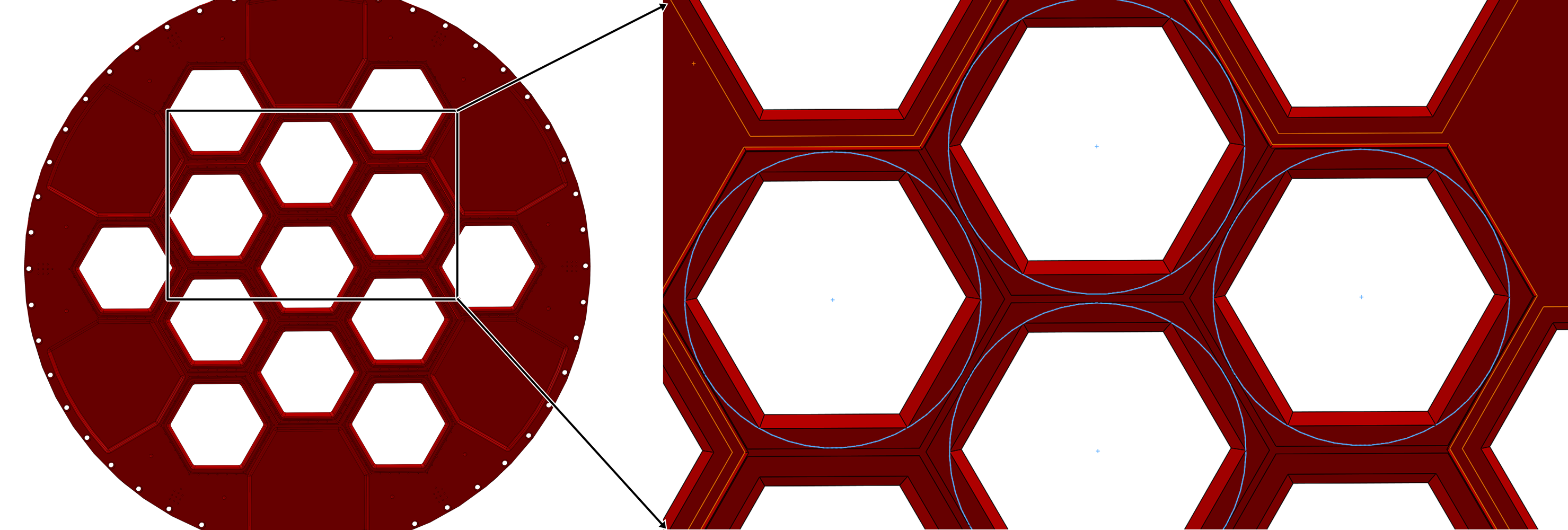}
   \end{tabular}
   \end{center}
   \caption[example] 
   { \label{fig:hexwindow} 
Closeup of front plate showing material saved by using hex windows. The gray circles are the size of the hole of equivalent minimum beam clearance at the front surface of the front plate. Making the windows inscribed hexagons instead of the circles which circumscribe them adds a significant amount of material at what is the weakest point in our front plate.}
   \end{figure}

The first challenge presented in designing the LATR is in the construction of the vacuum shell. While numerous examples of vacuum vessels as large or larger than ours exist, our design presented several specific challenges, the combination of which to our knowledge has not previously been solved. At the top of this list of challenges is the front plate design. The front plate of the cryostat needs to contain 13 densely packed apertures while still maintaining its structural strength. Further, the front plate needs to flex in less than 2~cm under load in order to avoid the 80~K stage behind it.\cite{Zhu2018} Finally, the plate has to be consistent with our weight restrictions--some early designs we considered weighed nearly 1000~kg, a significant portion of our total mass budget of 6000~kg.\cite{Zhu2018} The closest analog we have identified was the SPIDER cryostat.\cite{SPIDER} Their front plate, however, is highly curved and would exceed the length restriction on our cryostat. Our simulations identified the material surrounding the inner seven optics tubes as by far the most critical for the structural integrity of the cryostat. Therefore we endeavored to strengthen that area as much as possible. The plate is 1~cm thicker in this area in order to reduce stress concentrations. Further, each of our windows is a hexagon rather than the typical circle. Each optics tube projects a hexagonal optical beam which allows us to make our windows hexagonal to match. We thus remove less material as compared to a circumscribed circle, as shown in Fig.~\ref{fig:hexwindow}. We also taper the walls of each window the match the beam divergence, so that the bottom of our windows are slightly thicker than the tops of the windows. See Fig.~\ref{fig:Taper}. Since the front plate will be monolithic and will not require welding, we will manufacture it from 6061-T6 aluminum for optimum strength and machinability. Finally, we used FEA to identify areas of low stress and aggressively weight relieved them, resulting in the "pockets" visible on the front of the plate. The end result is a 7~cm thick front plate weighing 350~kg that deflects less than 1~cm while conforming to American Society of Mechanical Engineers (ASME) VIII-2 government.\\

\begin{figure} [ht]
   \begin{center}
   \begin{tabular}{c} 
   \includegraphics[width = .75\linewidth]{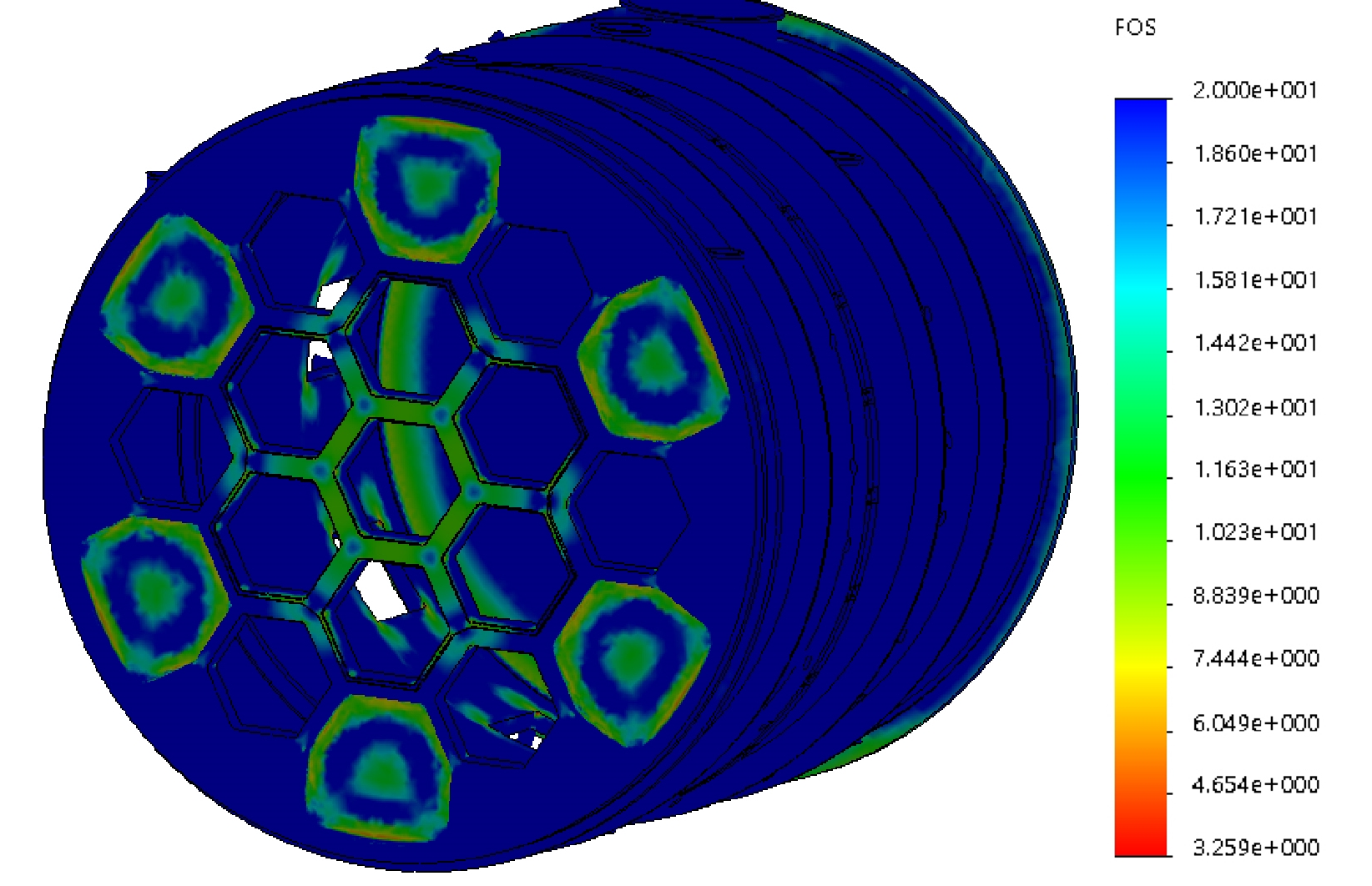}
   \end{tabular}
   \end{center}
   \caption[example] 
   { \label{fig:Vacuum FoS} 
Factor of Safety plot for the vacuum shell. The listed minimum of 3.2 is due to unphysical stress concentrations in corners. The actual minimum factor of safety on a surface is greater than 5 when fillets are accounted for. FEA allowed us to identify the correct shape for the weight relieving around the outside, as well as determine the maximum safe cut depth.}
   \end{figure}

The body of our vacuum shell follows a more conventional design. The body is split into two parts, a front shell and a back shell. This allows us easier access to the inside of the vacuum shell during assembly while also simplifying the manufacture of the shell. Along both halves are ribs placed roughly every 50~cm in compliance with the ASME VIII-1. The back half will be constructed from 0.25~in thick 6061 aluminum, while the front half will be 0.5~in thick. Our FEA, verified by an engineering firm specializing in pressure vessels called PVEng\footnote{Pressure Vessel Engineering Ltd, 120 Randall Dr b, Waterloo, ON N2V 1C6, Canada, https://pveng.com/}, indicates that 0.25~in is sufficient to resist the buckling mode. However the front plate flexing inwards stresses the front half of the shell. In order to resist this mode of failure, the front shell needs to be thicker. Our FEA validates 0.5~in as sufficiently thick. Finally, the back plate of the shell is a non-standard design which minimizes the distance from the flange of the back plate to the apex of curvature of the back plate. It was validated via FEA, as shown in Fig.~\ref{fig:Vacuum FoS}. The lowest factor of safety on our vacuum shell is approximately 5. Note that in Fig.~\ref{fig:Vacuum FoS} the minimum factor of safety is 3.2. This is a non-physical effect of the FEA, wherein stress concentrates in corners of the model much more than it would in reality. Meshing the fillets in the corners with sufficient density to remove the effect is computationally prohibitive. We find that the lowest factor of safety on a surface of our vacuum shell is at least 5.  \\

\subsection{80~K Stage}
\begin{figure} [ht]
   \begin{center}
   \begin{tabular}{c} 
   \includegraphics[width = .75\linewidth]{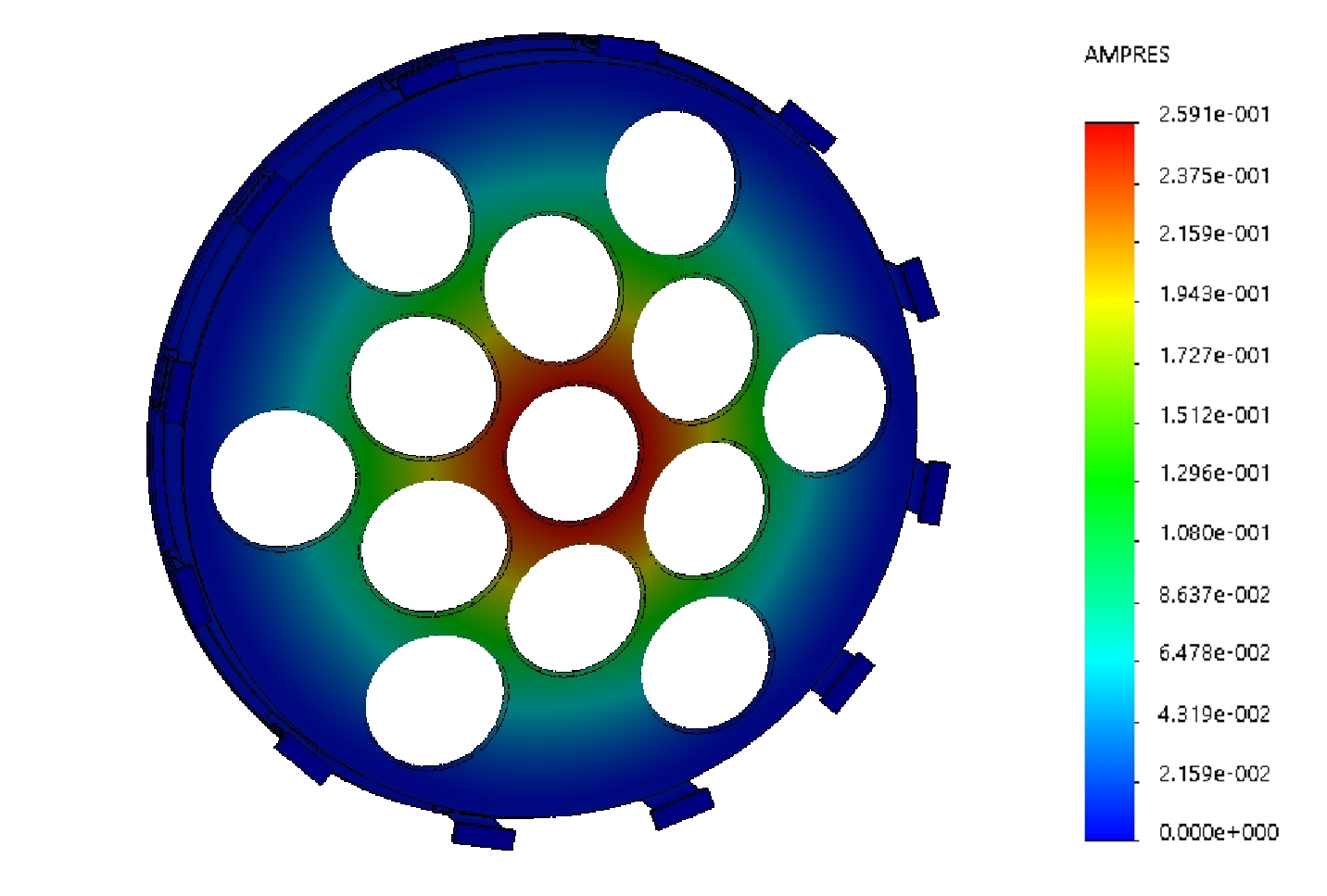}
   \end{tabular}
   \end{center}
   \caption[example] 
   { \label{fig:80K} 
Frequency analysis plot for 80~K filter plate. AMPRES is an arbitrary unit of resultant relative vibrational amplitude. The G10 tabs supporting the stage can be seen around the edge. The plot shows the drum-like vibration which is the first fundamental mode of this stage-the middle of the plate is flexing in and out.}
   \end{figure}

The 80~K stage supports only its own weight, and given the need for high thermal conductivity, we will construct the 80~K plate from 6063-T5 aluminum (see Sec.~\ref{sec:80therm}). The 80~K stage is suspended from the front vacuum plate, so this was selected as the fixture. To isolate the stage thermally from 300~K, we decided to construct the support structure from G10-CR, which is both thermally non-conductive and structurally strong. The results of our simulation were that the FoS was 62, the buckling load factor was 1700, and the first fundamental vibrational mode is at 46.9Hz as seen in Fig.~\ref{fig:80K}, all of which greatly exceed our requirements.

\subsection{40~K and 4~K Supports}
\begin{figure} [ht]
   \begin{center}
   \begin{tabular}{c} 
   \includegraphics[width = .75\linewidth]{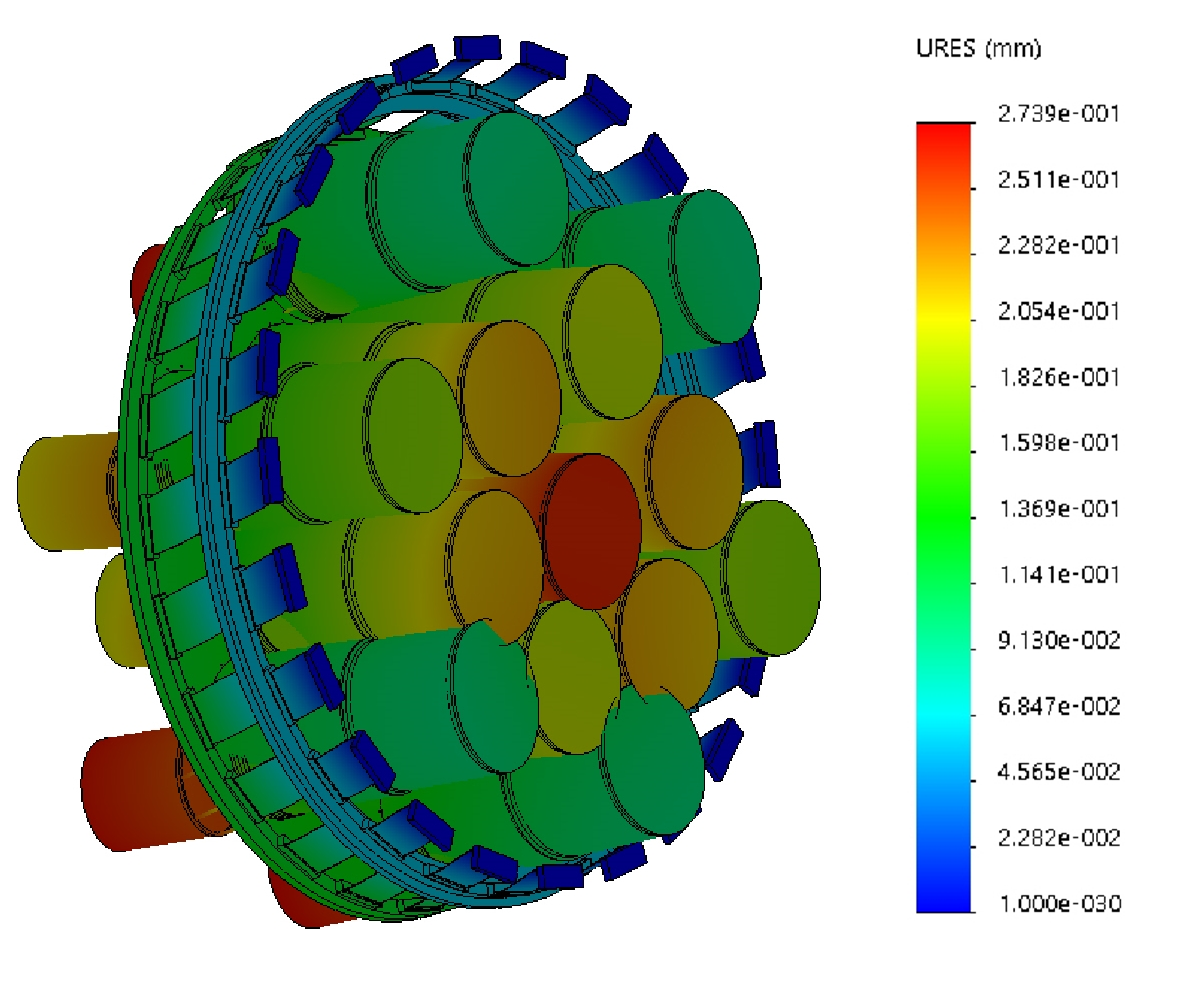}
   \end{tabular}
   \end{center}
   \caption[example] 
   { \label{fig:40K Disp} 
Displacement plot for the combined 40 and 4~K stages. The displacements show in the model are exaggerated to illustrate the directions of the displacements, and the optics tubes do not actually collide. The G10 tabs can be seen around the outside, with the 4~0K ring in between them. The 4~K plate is towards the left of the figure. The entire assembly is sagging towards the bottom of the figure, while the center 7 optics tubes are additionally sagging relative to the plate.}
   \end{figure}

We performed combined 40~K and 4~K simulations in order to include the structural coupling between the stages. Our fixture was the 300~K flange of the vacuum shell which supports the 40~K stage. For thermal isolation, both the 40~K and 4~K standoffs are constructed from G10-CRg. We find that the FoS is greater than 10, the load factor is 590, and the first vibrational mode is at 18.9~Hz. The FoS and load factor both meet our requirements, but the first vibrational mode is at a slightly lower frequency than desired. This mode involves the 4~K plate vibrating like a drum. We are currently investigating thickening the 4~K plate to raise this fundamental frequency. In addition to our typical requirements, we also have a sagging requirement on our optics tubes. Optics simulations show that the tip-tilt on the detector arrays at the back of the optics tube is 0.4 degrees to maintain our minimum Strehl ratio\cite{Dicker2018}, which corresponds to a maximum relative sag at the front of the optics tube relative to the back of the optics tube of 6~mm. We find that the tubes sag by approximately 0.3~mm, much lower than our requirement of 6~mm. Finally, due to the differential thermal contraction between the 300~K flange and the 40~K stage, the support tabs flex inwards radially approximately 6~mm. While we have performed FEA that shows that the tabs will be able to safely flex inwards by this amount, we have not been able to determine how to perform an analysis which includes both the mass load and the flexing inwards. As such, the 40~K tabs have been designed to have an extremely high FoS, exceeding 10, as seen in Fig.~\ref{fig:40K Disp}.\\

\subsection{Thermal Bus Support}
\begin{figure} [ht]
   \begin{center}
   \begin{tabular}{c} 
   \includegraphics[width = .75\linewidth]{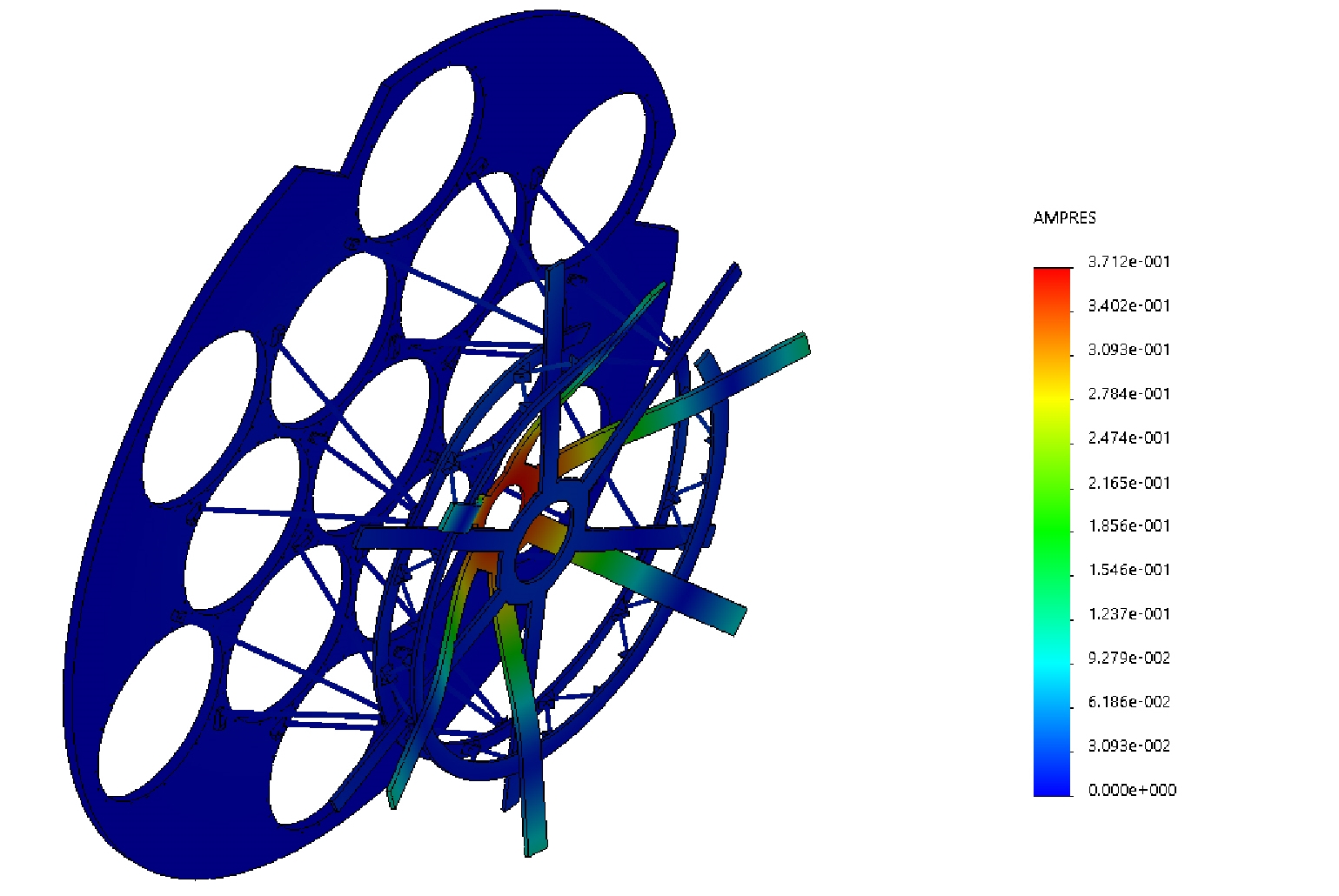}
   \end{tabular}
   \end{center}
   \caption[example] 
   { \label{fig:1K Freq} 
Frequency analysis plot for 1~K 100~mK thermal bus showing the first fundamental mode of vibration. This mode is at about 47~Hz and corresponds to a drum-like mode. AMPRES is an arbitrary unit of resultant relative vibrational amplitude. Since we do not include dampening, there is no way to calculate the amplitude of vibration at resonance. AMPRES shows the relative vibration of different parts of the assembly.}
   \end{figure}

The supports for our 1~K and 100~mK supports must be thermally isolative, like the other stages, in order to maintain the detectors and electronics at these stages at their operating temperatures. At these temperatures carbon fiber has very low thermal conductivity, lower than G10. In order to further reduce thermal conduction without sacrificing stiffness, the carbon fiber tubes are hollow. We simulated the 1~K and 100~mK supports together, fixing the 4~K plate to which the 1~K supports attach. We found that the current design meets our requirements, with a FoS of 89 and a buckling load factor of 58. Vibrational modes at these stages are particularly important, as vibration can cause appreciable heating and work harden the extensive amount of copper. We find that the first fundamental mode is at 28.6~Hz as seen in Fig.~\ref{fig:1K Freq}, above our requirement.

\section{THERMAL SIMULATIONS}
\label{sec:therm}
In addition to structural simulations, we have also completed numerous thermal simulations in order to quantify thermal gradients over various components in our receiver. The performance of many of the pieces of our receiver depends on the temperature of those components. In particular, the efficiency of our detectors, electronics, and optical elements depends on their temperatures. While we know the temperature of each stage's corresponding pulse tube or dilution refrigerator stage, there are frequently significant distances between the cooling apparatus and the elements they are cooling. We have designed our cryostat to minimize these distances, but in an instrument of this size, there will inevitably be longer thermal paths than in previous experiments, leading to correspondingly larger thermal gradients. Therefore, it is imperative that we develop an estimate for these thermal gradients, so that we can ensure that the optics, detectors, and readout electronics will be sufficiently cooled to function effectively.\\

The starting point for our thermal simulations is the Simons Observatory LATR thermal model. The model combines estimates of conductive loading through supports and wiring from one stage to another, radiative loading from warm stages to another, dissipation from electronics, and a simulation of the optical chain. To compute the conductive loading for a given part, we combine the cross sectional area A, length l, warm-side and cold-side temperatures, and a temperature-dependent conductivity model from the literature\cite{Woodcraft2009}\cite{NISTCryo} with the integral form of Fourier's Law:
\begin{equation}
\text{P} = -\frac{\text{A}}{\text{l}}\int_{\mathrm{t_{low}}}^{\mathrm{t_{high}}} \text{k(t)dt}
\end{equation}
For the radiative loading between stages, we use the Stefan-Boltzmann law, given in Eq,~2, conservatively assuming aluminum surfaces have a emissivity of $5\%$\cite{Bartl2004} and surfaces covered in multilayer insulation (MLI) have an emissivity of $.2\%$.\cite{Ross2015}
\begin{equation}
\text{P} = \text{A}\epsilon\sigma(\text{T}_{\text{high}}^{4}-\text{T}_{\text{low}}^{4})
\end{equation}
A is the absorbing area, $\epsilon$ is the emissivity, $\sigma$ is the Stefan-Boltzmann constant, and T$_{\text{low}}$ and T$_{\text{high}}$ are the colder and warmer stage temperatures. We also estimate and include the power dissipated by our readout electronics. To estimate the power deposited on each stage by the optical elements at that stage, we developed a thermal-optical model, which is described in a separate paper in these proceedings.\cite{Zhu2018} \\ 

We then use the COMSOL\footnote{COMSOL, Inc., 100 District Avenue, Burlington, MA 01803, USA, https://www.comsol.com} software suite to estimate thermal gradients across our 80~K, 40~K, 4~K, 1~K, and 100~mK stages. COMSOL is an FEA program that supports the simulation of various physical processes, including heat transfer. For a typical thermal simulation, We first import a computer aided design (CAD) model of our receiver into COMSOL. We then apply a material to each component, which sets its temperature-dependent thermal conductivity from models we have collected from the literature\cite{Woodcraft2009,NISTCryo}. We then apply the loads we calculated with our thermal model to the most appropriate locations on the model and fix the surface corresponding to that stage's cooling system to its operating temperature. Currently we use the manufacturer's guaranteed cooling power for our target temperature, but when we take delivery of the coolers we plan to measure their cooling curves, that is their cooling power as a function of temperature. From the loading on a given stage, we will compute the temperature that the refrigerator will reach, allowing us to refine our simulations. From there we mesh our model and run the simulation. Critically, our model does not take into account thermal resistance between material interfaces, which we call thermal joint resistance. COMSOL does have a function for computing this resistance, but we have not been able to verify its accuracy. Instead, we will make measurements of the thermal joint resistance over the most critical of our thermal interfaces in a test environment and combine these gradients with the ones computed in COMSOL to form a revised estimate of the total gradient. We then use the results of the simulations to inform our design decisions. For example, an earlier design for our 1~K and 100~mK thermal buses was much larger, nearly the size of our 4~K plate with only small cutouts. Via our thermal analysis, we were able to identify which regions were critical for conduction and which were not. We then removed material in the non-critical regions to arrive at our current design, which is significantly lighter and simpler to manufacture. \\ 

\subsection{80~K Plate}
\label{sec:80therm}
\begin{figure} [ht]
   \begin{center}
   \begin{tabular}{c} 
   \includegraphics[width = .75\linewidth]{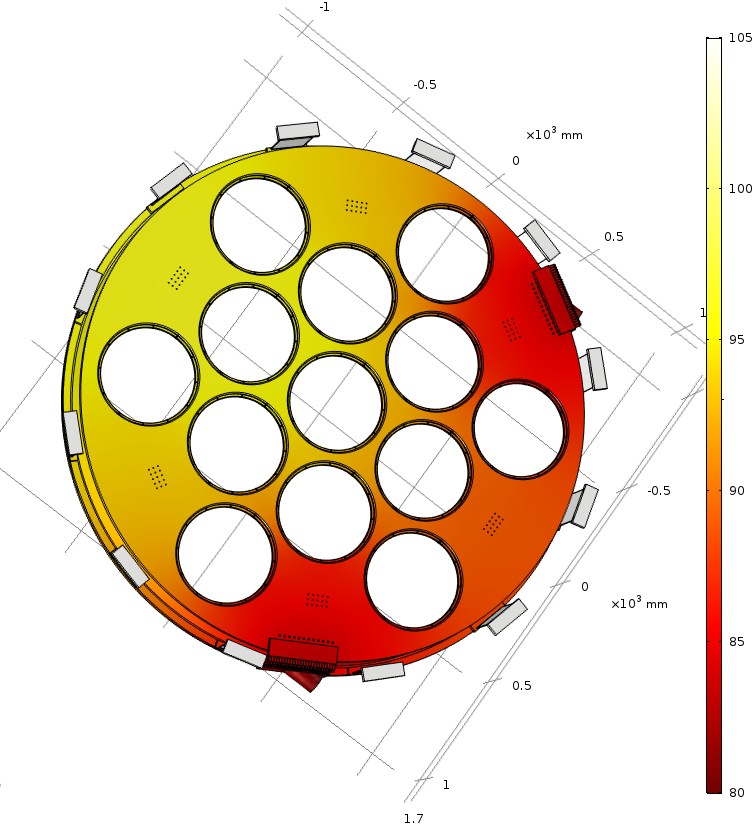}
   \end{tabular}
   \end{center}
   \caption[example] 
   { \label{fig:80K Thermal} 
Predicted thermal thermal gradient on 80K stage for 6063-T5 filter plate. The PT-90 pulse tubes would be attached to the thermal straps located on the right of the figure at the coldest points.}
   \end{figure}

The LATR will feature 80~K stage infrared (IR) filters that will be used to reduce the thermal loading on colder stages. These filters will consist of a double sided IR blocker and an alumina absorber.\cite{Zhu2018} We have determined that if the alumina filters are actually 120~K then our sensitivity will be reduced by less than $0.1\%$ as compared to the being at 80~K, which is an acceptable level\cite{Hill2018}. From our optical simulations, we expect the center of the alumina filters to be no more than a few Kelvin warmer than the edges.\cite{Zhu2018} Once we obtain these filters we will perform several tests, one of which will be to measure their thermal gradients. For now we conservatively estimate the gradient between the center of the filter and the edge as 20~K, so that our goal is to keep the gradient across the 80~K plate under 20~K. From our simulations, we find that a plate made of 6061-T6 aluminum would have a gradient of 27~K while a plate made of 6063-T5 aluminum will have a gradient of 16~K. Since we need further overhead reserved for thermal joint resistance, we therefore have decided to construct our 80~K plate from 6063-T5 aluminum. See Fig.~\ref{fig:80K Thermal}.

\subsection{40~K Stage}
\begin{figure} [ht]
   \begin{center}
   \begin{tabular}{c} 
    \includegraphics[width = .75\linewidth]{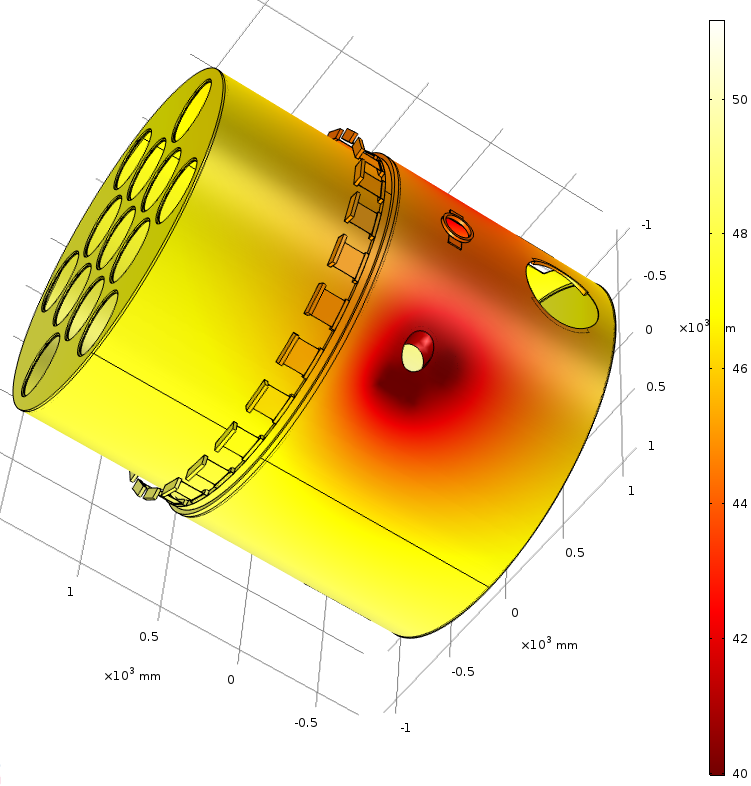}
   \end{tabular}
   \end{center}
   \caption[example] 
   { \label{fig:40K Thermal} 
Predicted thermal thermal gradient on 40K stage for 6063-T5 extension tube, radiation shield, and filter plate. The pulse tubes are not shown but connect to the coldest points. Note that even with 6063-T5, the filters are nearly at 50~K at their edges.}
   \end{figure}

The 40~K stage presents a unique challenge in that the sources of thermal loading are far away from the cryo attachment points. Our current design calls for a 40~K structural ring, supported from 300~K and supporting 4~K, to which also are attached a 40~K radiation shield, a standoff, and 40~K filter plate.\cite{Zhu2018} These shields and the filter plate absorb most of the power on the 40~K stage. Therefore, we would ideally make the shield thick to reduce thermal gradients along the 40~K stage to the filter plates and readout. However, to meet our mass budget, these shields must be rather thin, at 0.125~in. Our FEA shows that making shields of this size with our predicted loading out of 6061-T6 aluminum would result in unacceptably high gradients of 25~K. Therefore, we will construct the 40~K shield and extension tube from 6063-T5 aluminum, which offers nearly an order of magnitude more thermal conductivity than 6061-T6 at 40~K\cite{Woodcraft2010} while still maintaining more than half the strength.\footnote{From Solidworks Simulation materials library} \setcounter{footnote}{0} We will still make the 40~K structural ring out of 6061-T6 aluminum. This combination of materials allows us to maintain a factor of safety in excess of 4 over the 40~K stage, while achieving a thermal gradient of only 7~K, ensuring that our filters and electronics operate as expected. See Fig.~\ref{fig:40K Thermal}. \\

\subsection{4~K and Optics Tubes}
\begin{figure} [ht]
   \begin{center}
   \begin{tabular}{c} 
   \includegraphics[width = .75\linewidth]{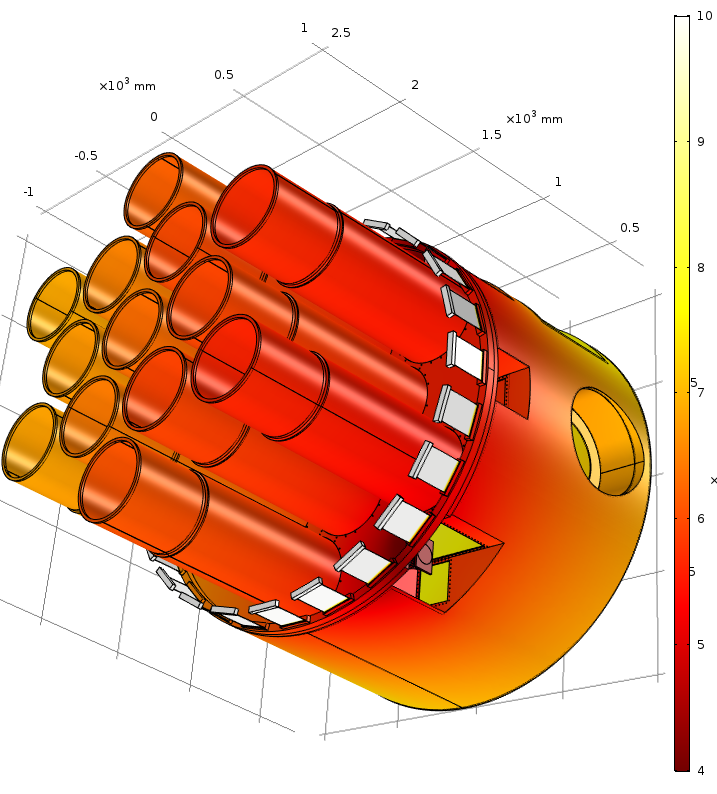}
   \end{tabular}
   \end{center}
   \caption[example] 
   { \label{fig:4K Thermal} 
Predicted thermal thermal gradient on 4~K stage for 1100-H14 optics tubes. PT 420's are not shown but attach to the thermal straps, one of which can be seen at the middle, at the coldest point. Note that the gradients on the radiation shield are relatively unimportant--while it increases the radiative loading on the stages inside the shield, there is no other performance degradation.}
   \end{figure}

For the 4~K stage, we quickly determined from structural FEA that the plate would have to be made of 6061-T6 aluminum. Our structural analysis also determined that the deflection of the optics tubes would be acceptable if they were made of either 6061-T6 or 1100-H14 aluminum, though the deflection is somewhat worse with 1100-H14. We therefore analyzed both materials in our thermal FEA. Our simulations show that the total gradient for a 6061-T6 aluminum plate coupled to 6061-T6 optics tubes would be 5.2~K and for a 6061-T6 aluminum plate coupled to 1100-H14 aluminum optics tubes would be 2.9~K (see Fig.~\ref{fig:4K Thermal}). Since the 6061-T6 aluminum gives only marginal gains in tube sag, we will construct our tubes out of 1100-H14 aluminum. Since we rely on our flanges to hold the shape of the optics tube, we will construct the flanges from 6061-T6 aluminum and then glue the 1100-H14 aluminum tubes into the flanges. For these interfaces we are investigating the use of a cryogenic epoxy designed for good heat conduction\footnote{Epotek T7110. See https://www.epotek.com}. This epoxy is weaker than the epoxy we use for more structurally critical components, like the 40~K support tabs, but still strong enough for the optics tubes. When we take reception of our first optics tube, we plan on testing both the strength of the epoxy and its thermal conductivity. This will allow us to precisely evaluate the thermal joint resistance along much of our 4~K thermal path, refining our thermal gradient estimate. If we find the thermal gradients are excessive, then we will attach strips of 4N high purity aluminum to the outside of our optics tubes to reduce their gradients. 

\subsection{Thermal Bus}
\begin{figure} [ht]
   \begin{center}
   \begin{tabular}{c} 
   \includegraphics[width = .75\linewidth]{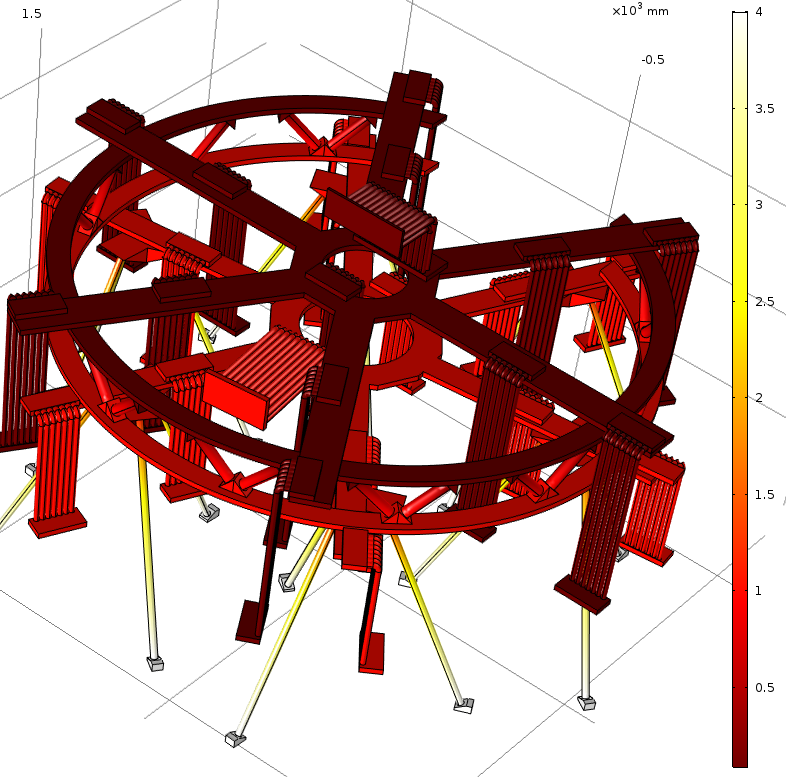}
   \end{tabular}
   \end{center}
   \caption[example] 
   { \label{fig:1K 100mK Thermal} 
Predicted gradient on the 1~K-100~mK thermal bus. The 100~mK bus is on top, with the 1~K bus below. The thermal bus distributes power to the detector arrays and optical elements at 1K, so minimizing the gradients across the bus is critical. Our current best simulations show a gradient of about 2~mK across the 1~K stage and 8~mK across the 100~mK stage.}
   \end{figure}

To conduct heat away from our detectors to our dilution refrigerator (DR), we employ two thermal buses constructed of oxygen free high conductivity (OFHC) copper, one for the 1~K stage and one for the 100~mK stage. Minimizing the gradients on these stages is particularly important, as the sensitivity of our instrument is strongly dependent on the temperature of the Lyot stop, which is nominally at 1~K, and of the detector arrays, which are nominally at 100~mK. We therefore elected to make the buses out of OFHC copper. Since OFHC copper is heavy and expensive, we worked to minimize the size of the bus while retaining adequate performance thermally. We started with an oversized design, used FEA to identify areas that were not conducting large amounts of heat, and removed those areas to arrive at our current design. This design also has the property that it is relatively easy to manufacture and work with. The arms and core can be machined separately and then welded together, while the spacing between the arms provides ample room to reach through and work on interfaces behind it. We also included rough designs for heat straps in our model to achieve the most realistic results. Excluding thermal joint resistance, we found that the maximum gradient over the 1K thermal bus and straps is 0.6~mK and the maximum gradient over the 100~mK stage and straps is 4.4~mK (Fig.~\ref{fig:1K 100mK Thermal}). While the loading on the 1~K stage is higher than on the 100~mK stage, the thermal conductivity of OFHC copper is also much higher, resulting in the 1~K gradients being smaller than the 100~mK gradients. While have not yet received our DR to make cooling curves, we do have cooling curves from a similar, but less powerful DR from the same manufacturer which we used as an estimate of the 1~K and 100~mK base temperatures. We took the base temperatures to be are 1~K and 80~mK respectively, so that the warmest temperature predicted on the 1~K stage is 1.0006~K and the warmest on the 100~mK is 84.4~mK. We do not yet have a model for the detector arrays, and thus we do not yet have a model of their gradients. However, the thermal loading on the detectors is much lower than other loads on the 100~mK stages, so the gradient across the detector arrays should be small compared to the gradients across the bus to the array. In general, from our experience on previous experiments,\cite{2016ApJS..227...21T} we expect that thermal joint resistance will dominate at these temperatures. As such the simulations above constitute a lower bound on the expected thermal gradient at these stages. Therefore, we plan to weld each braid to the thermal bus, and to weld each braid between our detector arrays and the 100~mK thermal bus at the thermal couple coming from the detector array. We will also measure these thermal joint gradients during cryogenic acceptance. \\

\section{SUMMARY}
Computer assisted design and FEA simulations have been key for the design of the Simons Observatory LATR. As we iterated through designs of various components, FEA provided critical feedback on the performance of those components. This feedback allowed us to evaluate the performance of our design, determine what could be improved, and design the next iteration. Through this process, we were able to converge on our current and final design for the LATR. Lastly, FEA provided the final validation of each component, ensuring for us that they would meet the performance specifications that we set. The result of this process is a design which is validated and slated for manufacture in the near future, and which can inform the design of a future CMB-S4 instrument.\cite{Abitbol2017}\cite{Abazajian2016}

\appendix    

\acknowledgments 
 
The LATR team and the SO collaboration would like to thank the CCAT-prime team for allowing us to borrow heavily from their telescope design. We thank the Vertex group who have been working closely with us on the telescope development.  Meyer Tool\footnote{Meyer Tool and Mfg., 4601 SW Hwy, Oak Lawn, IL 60453, USA, https://www.mtm-inc.com/}, Dynavac\footnote{Dynavac, 10 Industrial Park Rd \# 2, Hingham, MA 02043, USA, https://www.dynavac.com/}, PVEng, and Fermilab have provided a number of helpful suggestions regarding the cryostat design and fabrication process. Bluefors\footnote{Arinatie 10, 00370 Helsinki, Finland, https://www.bluefors.com}, Cryomech\footnote{113 Falso Dr, Syracuse, NY 13211, https://www.cryomech.com},  Oxford Instruments\footnote{Oxford Instruments, Tubney Woods, Abingdon, Oxfordshire, OX13 5QX, United Kingdom, https://www.oxford-instruments.com/}, and High Precision Devices\footnote{3241, 4601 Nautilus Ct S, Boulder, CO 80301, https://www.hpd-online.com/} generously provided detailed cooling capacity information on their refrigeration units. Finally, we would like to thank Solidworks\footnote{Dassault Syst\`emes, 10, Rue Marcel Dassault, 78140, V\'elizy-Villacoublay, FRANCE, https://www.solidworks.com/} and Comsol\cite{Comsol} for their excellent FEA software.

This work was supported in part by a grant from the Simons Foundation (Award \#457687, B.K.).

\bibliography{ref} 
\bibliographystyle{spiebib} 

\end{document}